# Defining Pathway Assembly and Exploring Its Applications

Alastair R. G. Murray, Stuart M. Marshall, Leroy Cronin*

School of Chemistry, University of Glasgow, Glasgow, G12 8QQ, UK. *Corresponding author email: Lee.Cronin@glasgow.ac.uk

**Abstract**

How do we estimate the probability of an abundant object's formation, with minimal context or minimal assumptions about its origin? To explore this we have previously introduced the concept of pathway assembly, in a graph-based context, as an approach to quantify the number of steps required to assemble an object based on a hypothetical history of an object's formation. By partitioning an object into its irreducible parts and counting the steps by which the object can be reassembled from those parts, and considering the probabilities of such steps, the probability that an abundance of identical such objects could form in the absence of biological or technologically driven processes can be estimated. Here we give a general definition of pathway assembly from first principles to cover a wide range of cases, and explore some of these cases and applications which exemplify the unique features of this approach.

**Introduction**

In a recent publication we defined pathway-assembly (referring to it as pathway complexity at the time, but we have since renamed it to avoid ambiguity, due to the widely varying contexts in which the notion of "complexity" is understood [1]). It is a new measure designed to track the likelihood that objects could have formed in the absence of biological processes. Its development was motivated by a desire to define a threshold on the length of a constructive process, above which biological processes must be involved, and to use it to detect life and to study the transition between non-living and living systems. This is because we wondered if such an approach to identify the common features of artefacts produced by biology might be more fruitful than trying to explore commonly accepted things about what life does, or searching for criteria that define life. By searching for artefacts that have too many features to have emerged without a 'rule book' we might change the nature of the exploration. As such, we can look for alien / inorganic life, or observe the assembly of life-like systems from abiotic building blocks that have formed or operate in the absence of a genetic system or external information-based read-control-assembly system.

In this paper we explore pathway assembly from a mathematical standpoint, giving a general but precise definition, and exploring its behaviour and bounds. Following this we compare pathway assembly to a number of other measures in common use, with particular reference to their relative utility in establishing a threshold. Finally, we explore a number of possible applications pathway assembly in different contexts.

## Pathway Assembly

Pathway assembly [2] was conceived primarily as a process-neutral approach to the formation of physical objects such as complex molecules, human-created artefacts, and other potential biosignatures, by considering the acquisition of processes, in steps, for the construction of the object without any external directed intervention. By allowing random processes to assemble the object in steps, it is possible to consider how likely it is that a given object can be produced by such processes if it is found in any abundance. By thresholding this we hope to use pathway assembly as a tool to distinguish randomly occurring objects from objects assembled via the biases common to biological processes. As such, pathway assembly seeks to represent hypothetical nontrivial histories of the objects considered within the assembly process. It has a recursive nature that makes use of repeatable symmetries of objects, and a simple context-independent constructive process analogous to the formation of physical objects. A timeline of the emergence of some of the key concepts described above is shown in Figure 1.

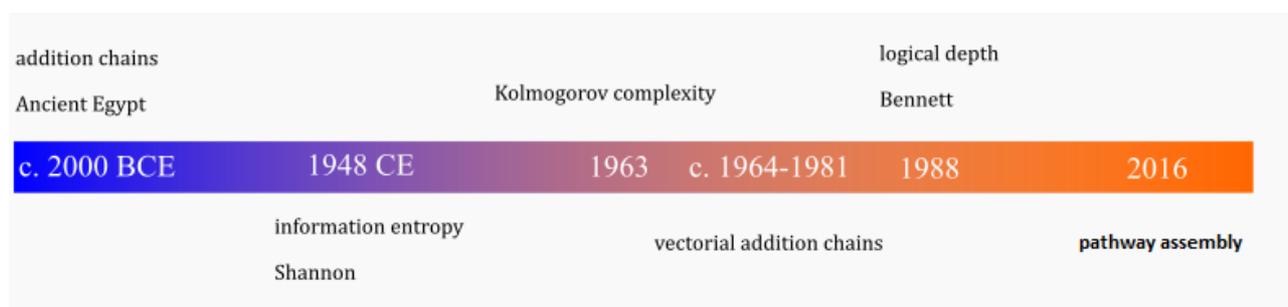

**Fig. 1.** A brief timeline of the development of some select ideas relevant to the foundations of pathway assembly.

## Definitions

The pathway assembly index of an object $X$ is defined with respect to two things: a set of basis objects (or units), and a composition rule. The set of basis objects is a set $B$ of objects that can be considered "fundamental" in the construction of further objects. Their choice is in principle arbitrary, but different contexts often naturally suggest a particular set of units. The composition rule is a function $f$ that maps any pair of objects $X$ and $Y$ to another set of objects – the objects that can be "built" from $X$ and $Y$ in one step.

To ensure that pathway assembly is computable (i.e. that a method to calculate it exists that is guaranteed to finish for any input), it is assumed that both the set $B$ and all sets in the range of $f$ are finite, and that it is known whether or not any given object has at least one valid assembly pathway. As demonstrated later, these two conditions ensure that a method always exists to systematically search the space of assembly pathways. Note that some definitions below – namely, content-preserving measures and unique decomposition – are defined very similarly to other well-established mathematical concepts.

**Definition 1.** An **assembly pathway** $S$ relative to $(B, f)$ is a sequence of objects $(x_{-m}, x_{-m+1}, \ldots, x_1, x_2, \ldots, x_n)$ such that:
- If $i < 1, x_i \in B$
- If $i \geq 1, x_i \in f(x_j, x_k)$ for some $j, k < i$

In other words, an assembly pathway is a sequence of objects in which every object is either a basis object, or can be formed by combining two previous objects in the sequence.

**Definition 2.** The **space of objects** $U$ of $(B, f)$ is the set of objects that can appear in assembly pathways constructed relative to $B$ and $f$. It is uniquely determined by the choice of $B$ and $f$ like so:
- $B \subseteq U$
- If $x, y \in U, f(x, y) \subseteq U$
- $U$ is the set defined iteratively by the above conditions.

In other words, the space U contains all objects that can be reached by repeated applications of $f$ to other objects, starting only from members of B.

**Definition 3.** The **pathway length** $l(S)$ of an assembly pathway $S$ relative to $(B, f)$ is $|S \backslash B|$, i.e. the number of non-unit objects in S.

$$l(S|B, f) = |S \backslash B|$$

**Definition 4 (main definition).** The **pathway assembly index** $c(X)$ of an object $X$ relative to $(B, f)$ is the minimum pathway length of all pathways $S$ that contain $X$, i.e.

$$c(X|B, f) = \min(\{l(S|B, f) \ \forall \ S \ | \ X \in S\})$$

We write simply $c(X)$ if $B$ and $f$ are implied.

Features of assembly pathways, and the choices involved at each stage of their construction, are shown in figures 2 and 3.

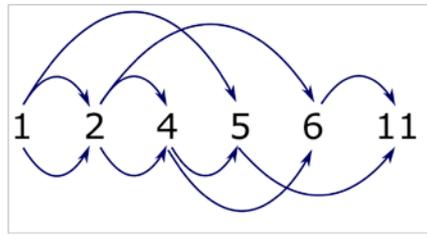

**Fig. 2.** An assembly pathway for which $B = \{1\}$ and $f(x,y) = \{x+y\} \, \forall x, y$. (This is identical to the addition chains discussed later.) Every step of the pathway (bar the first, which is the only basis object) is obtained from the two previous steps whose arrows lead to it. Note that this pathway could also be built via different choices: 6 could be obtained from 1 and 5 instead of from 2 and 4.

**Fig. 3.** Features of an assembly pathway being constructed. From a pool of objects (first column), two are selected (second column); from all possible combinations of those two objects (third column), one is added to the pathway (fourth column), and is now available for future constructions. In this example, red and green squares are the basis objects, and they can be combined edge-to-edge, where different orientations are considered distinct. The second-last step is redundant, though extensions of the pathway could potentially make use of it.

**Definition 5.** The **pathway co-assembly index** $cc(\{Xi\})$ of a set of objects $\{Xi\}$ relative to $(B, f)$ is the minimum pathway length of all pathways S that contain all members of $\{Xi\}$, i.e.

$$cc(\{Xi\}) = \min(\{l(S|B,f) \; \forall \; S \mid Xi \in S \forall Xi \in \{Xi\}\})$$

Pathway co-assembly is a useful supplemental concept, both for considering the concurrent construction of multiple target objects, and in the construction of algorithms to compare assembly pathways.

**Definition 6.** A **content-preserving measure** defined with respect to a function f is a function $m: U \to \mathbb{R}$ such that $\forall X, Y, Z$, if $X \in f(Y, Z)$, then $m(X) = m(Y) + m(Z)$.

In other words, any property of $Y$ and $Z$ that is represented by the measure is "conserved" when the two objects are combined.

**Definition 7.** A set of objects is **uniquely decomposable** with respect to $(B, f)$ if $\forall b \in B \; \exists$ a content-preserving measure $m_b$ with respect to $(B, f)$ such that:

- $m_b(b) = 1$
- $m_b(x) = 0 \; \forall x \in B \backslash \{b\}$

In the most intuitive interpretation, the measure $m_b$ counts how many of the unit b appear in an object, and this quantity is consistent no matter what assembly pathway is used to construct the object.

**Definition 8.** The **size** of a uniquely decomposable object $X$ is $m(X)$ for the $m$ uniquely defined by $m(b) = 1 \; \forall b \in B$.

In other words, all units are considered to have a size of 1, and the size of all subsequent objects is found by adding together the sizes of its components – a process that can be iterated until the original units are reached.

## Results and Discussion

### The behaviour and limits of the measure

**Proposition 1.** A content-preserving measure is uniquely determined by its value for basis objects.
**Proof:** by induction.

**Proposition 2.** $c(b) = 0 \ \forall b \in B \ \forall \ (B, f)$

i.e. all basis objects have a pathway assembly index of 0.

**Proof.** The length 1 sequence $(b)$ is a valid assembly pathway for the unit $b$, and as the pathway must contain $b$ itself, there are no shorter pathways. So,

$c(b) = length((b)) = 1 - 1 = 0$.

**Proposition 3.** If at least one assembly pathway containing an object $X$ is known to exist, its pathway assembly index is computable.

**Proof.** As the number of units in $B$ is finite, and $f(x)$ is finite $\forall x$, the space of pathways can be systematically searched in size order until a pathway is found. The known pathway will be found if no other pathway is found before it.

## Upper and lower bounds

A few basic deductions provide a useful starting point for finding upper and lower bounds on the pathway assembly index of an object.

**Proposition 4.** If an object has a unique decomposition, and thus a size $N$, then $N - 1$ is an upper bound for the pathway assembly index.

**Proof.** By induction. In the base case, all objects of size 1 are basis objects of pathway assembly index 0. Suppose all objects of size $N$ or less have pathway assembly indices of at most $N - 1$. Then every object $Z$ of size $N + 1$ must be a combination of two objects $X$ and $Y$ of sizes $M$ and $N + 1 - M$ respectively, for some $0 < M < N + 1$. An assembly pathway for the object of size $N + 1$ can be constructed by stating assembly pathways for $X$ and $Y$ sequentially (omitting the basis objects the second time), followed by a final step that combines $X$ and $Y$ to make $Z$. The pathway lengths are at most $M - 1$ for $X$ and at most $N + 1 - M$ for $Y$, and so Z has an assembly pathway for which the pathway length is at most $M - 1 + N + 1 - M + 1 = N + 1$.

When constructing algorithms, general lower bounds are more important to find than general upper bounds, as the length of any valid assembly pathway for an object X is itself an upper bound for the pathway assembly index of X. Later in this paper, we discuss a "tree" algorithm that considers some, but not all, assembly pathways that construct a uniquely-decomposable object, and finds the shortest such pathway; these provide upper bounds on the pathway assembly index. Some established complexity measures also serve as bounds for pathway assembly indices. These are described later in the paper in the section on comparisons with complexity measures.

**Correlations**

In circumstances where the number of possible pathways of specified lengths grows quickly, and where the objects under consideration are randomly generated, size is the clearest correlate with pathway assembly index. In objects such as graphs or text strings, randomly generated objects built from a sufficiently large set of basis units or with many possible means of combining components, few repeated structures appear, and so the pathway assembly index is very close to the object's size.

When the composition rules are based on geometric combinations, highly symmetric objects, and objects with many symmetric sub-objects, tend to have a much lower pathway assembly index than asymmetric ones, as they contain (often nested) repeated components. The objects with the highest ratio of size to pathway assembly index are the ones where every object in the pathway's sequence is formed by combining two copies of the previous object, except the first object which is a unit; such objects of assembly index n have size $2^n$. However, depending which symmetries affect the permitted compositions, objects such as those in figure 4 can have varying symmetry but identical assembly indices.

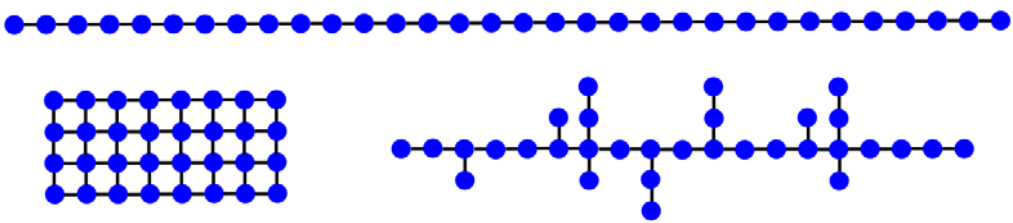

**Fig. 4.** Non-symmetries allowed by the composition rule can result in qualitatively different objects having similar pathway assembly indices. Under our standard formulation of the pathway assembly of graphs, these three objects all have a pathway assembly index of 5, as adding edges during compositions is "free". (For all three objects, at each step, two copies of the same object are added together.)

A small breaking of the symmetry of an object may have a large effect on its pathway assembly index. For example, in uniquely decomposable objects (where by necessity two objects can only be added together, without being taken apart), introducing a "hole" into an object by removing a small component may mean that, instead of being able to duplicate a large component in one step, it may take several steps to build around the hole (or, alternatively, the fastest pathway may be to make the "broken" component, and then patch it up to the original one). For example, the shortest addition chain (as defined elsewhere) for the number 128 has 7 steps, whereas the shortest addition chain for 127 (effectively 128 with a "hole" of size 1) has 10 steps. If the composition rule allows for deletion of components, this particular phenomenon may be far less pronounced.

**Assigning probabilities to pathways**

For any set of pathways for a universe of objects, abstractions of physical processes could be described that allow probabilities to be assigned to each pathway, and from that to the objects themselves. In this regard we can consider the set of all assembly pathways defined with respect to some $(B, f)$, and a probability distribution that, given a pathway of length $n$, assigns a probability to all pathways of length $n + 1$. Since any arbitrary probability distribution can be assigned to the pathways, it is impossible to make any *a priori* conclusions about certain pathways being "unlikely" without knowledge of them. Consider a "favoured" sequence of pathways $(S_1, S_2, S_2, ...)$ such that each pathway after $S_1$ is an extension of the pathway before it. As an example, each member of the sequence could be assigned a probability of 1 (given that a pathway of its length forms), and all other pathways of the same length could be assigned a probability of 0. This would leave the combinatorial explosion of pathways at larger lengths utterly irrelevant. Nor is it enough to insist that all pathways have a non-zero probability: in this case, each $S_n$ could be assigned a probability of ½ $+ \frac{1}{2^n}$. $P(S_n) >$ ½ $\forall n$, and as $P(Sn + 1) > P(S_n) \forall n$, all other pathways can be assigned non-zero probabilities.

However, mild conditions on the permitted probability distributions result in small probabilities for all sufficiently long pathways. For example, this condition suffices: if a pathway of length n can be extended into more than one pathway of length $n + 1$, then the probability that a pathway of length $n + 1$ is created given that the preceding pathway of length n is created cannot exceed some probability $q$. Thus, any pathway of length $n$ has a probability of at most $q^n$ of being created. This does not counter the possibility that multiple pathways may lead to the same object.

Of course, there are any number of reasons why a model of a physical system might have "dominant" pathways. However, this fact may demand explanation. If a system which would naïvely be assumed to select pathways with equal (or near-equal) probabilities instead disproportionately selects specific pathways, or classes of pathways, then some process beyond random selection is implied. This might be as simple as large classes of pathways being explicitly impossible, something more involved such as a "simple" physical step that takes many steps to represent in an assembly pathway, or else be subject to a nontrivial physical process.

## Algorithms

If the objects under consideration have some implementable representation, assembly pathways can be stored as lists of objects. A naïve and generally inefficient algorithm for finding the pathway assembly index of an object X is as follows:

> 1. Store a pathway that contains only units (do not count those objects towards the pathway length).
> 2. Set n = 0.
> 3. Until a pathway containing X has been found:
>    For every stored pathway S of length n:
>    For every pair of objects Y and Z in P:
>    For every possible composition of Y and Z:
>    Store a pathway consisting of S with the new composition appended to it.
>    Increment n by 1.
> 4. The pathway assembly index of X is the pathway length of the pathway that contains X.

We shall assume for the remaining discussion of algorithms that the choice of $B$ and $f$ for the pathway assembly measure being used has unique decomposition for all its objects, which yields more easily to the methods we have attempted. In many contexts, it can be checked whether an object on a pathway is unusable. For example, if an object is uniquely decomposable, then only steps that create its sub-objects need be considered. By tracking the quantity of basis objects (or combinations thereof) in objects in a pathway, the choices can be further narrowed down, as described in the algorithm below (in which the powerset of a set is a collection of all its subsets).

> 1. Find all sub-objects of X that appear more than once in the object.
> 2. Find a viable pathway for X; its length provides an upper bound on the assembly index. The size of the object is the simplest such bound.
> 3. Find vectorial addition chains that correspond to the object, where each component is a content-preserving measure relative to (B, f) (a 1-dimensional addition chain may be sufficient, or the most computationally cost-effective). The shortest such chain is a lower bound on the assembly index.
> 4. Create a set of pathways whose members are taken from the powerset of subobjects of X. Only consider pathways with length less than the established bound.
> 5. The assembly index of X is the length of the shortest pathway that contains X.

In some circumstances, it is possible to make a logical deduction that two distinct sub-objects cannot both be constructed on an assembly pathway for an object without the pathway being suboptimal, again allowing options to be narrowed. We can consider also a pragmatic "tree" algorithm where not all pathways are considered, but which allows for an easier systematic search of considered pathways.

Tree algorithm

> 1. Find all sub-objects of X that appear more than once in the object.
> 2. Find all partitions of X into repeating sub-objects and units.
> 3. Find the assembly index of each partition, using this algorithm recursively on sub-objects as necessary.
> 4. If the assembly index of the object by a certain partition is determined to be above an established threshold, disregard that partition.

## Non-exhaustive algorithms

We are working towards developing a greedy algorithm for which local optimisation of pathways is efficient, though many approaches are inefficient. However, in some tested cases, large numbers of pathways can be randomly generated in a short amount of time, and several of these are often close in size to the shortest pathway.

## Comparison with other measures

In this section we describe other measures related to pathway and compare them to it.

## Addition Chains

Addition chains, and their multi-dimensional generalisation vectorial addition chains, are special cases of assembly pathways. They have provided a conceptual foundation for our own definitions. An addition chain has been defined [3] as "a finite sequence of positive integers $1 = a0 \leq a1 \ldots \leq ar = n$ with the property that for all $i > 0$ there exists a $j, k$ with $ai = aj + ak$ and $r \geq i > j \geq k \geq 0$. An optimal addition chain is one of shortest possible length $r$ denoted $l(n)$." Efficient algorithms have been designed to calculate optimal addition chains; in [3], for example, optimal chains for numbers up to $2^{232}$ were systematically calculated. General results on the bounds of addition chains have been found, and other properties such as the Scholz–Brauer conjecture $l(2n - 1) = l(n) + n - 1$ have been investigated, and often partially confirmed. Vectorial addition chains are a generalisation of addition chains, used as a method of calculating monomials [4]. They have not been as thoroughly studied as addition chains, but papers such as [5] describe the investigation of their properties.

Addition chains are a special case of assembly pathways within the following framework:
- $B = \{1\}$
- $f(a,b) = \{a+b\} \forall a, b \in \mathbb{N}$
- $U = \mathbb{N}$, the set of natural numbers

Thus, as the definition of the shortest assembly pathway and the shortest addition chain are here equivalent, the shortest addition chains for any number are the shortest assembly pathways for the same numbers. Addition chains serve a useful purpose as a simple but non-trivial example of assembly pathways, as they have already been well-studied, their exponential growth is slower than many other examples, and they provide a lower bound for the lengths of other types of assembly pathway.

Vectorial addition chains, similarly, are also assembly pathways, as follows:
- $B = \{(1, 0, \ldots, 0), (0, 1, \ldots, 0), \ldots, (0, 0, \ldots, 1)\}$, the standard basis of unit vectors for an n-dimensional vector space
- $f(a,b) = \{a+b\} \forall a, b \in \mathbb{N}^n$
- $U = \mathbb{N}^n \setminus \mathbf{0}$, the space of n-dimensional vectors of natural numbers, excluding the $\mathbf{0}$ vector.

**Proposition 4.** Consider a set $B$ of basis objects, a composition rule $f$ under which all objects built with respect to $(B, f)$ are uniquely decomposable, a content-preserving measure $m$ defined with respect to $f$ such that $m(b) \in \{0,1\} \forall b \in B$, and an assembly pathway $(x_1, x_2, \ldots, x_n)$ for an object $x_n$ defined with respect to $B$ and $f$. The sequence $(m(x_1), m(x_2), \ldots, m(x_n))$ is an addition chain for $m(x_n)$.

**Proof.** By induction: An assembly pathway consisting only of basis units trivially has this property. Consider an assembly pathway $S$ that has this property, giving a corresponding addition chain A. Extend it to an assembly pathway S' by appending an object $Z$ to $S$, such that $Z \in f(X, Y)$. As $m(Z) = m(X) + m(Y)$, the corresponding sequence of numbers has the properties of an addition chain (perhaps with some redundant steps).

**Lemma 5.** The shortest addition chain for the size of a uniquely-decomposable object X is a lower bound on the pathway assembly index of X.

**Proof.** Let $n$ be the length of such an addition chain. All assembly pathways for $X$ have corresponding addition chains of the same length as the assembly pathway. Hence, all assembly pathways for $X$ must have length of at least $n$, or else there would be a corresponding shorter addition chain with length less than $n$.

Similarly, vectorial addition chains based on multiple measures provide lower bounds for the pathway assembly index of the object. Figure 5 demonstrates how corresponding assembly pathways, vectorial addition chains and addition chains can be constructed from and compared to each other.

| step | assembly pathway | vectorial addition chain | addition chain |
|---|---|---|---|
|  |  | (1, 0, 0) | 1 |
|  |  | (0, 1, 0) |  |
|  |  | (0, 0, 1) |  |
| 1. | 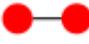 | (2, 0, 0) | 2 |
| 2. | 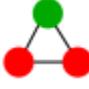 | (2, 1, 0) | 3 |
| 3. | 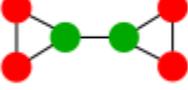 | (4, 2, 0) | 6 |
| 4. | 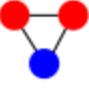 | (2, 0, 1) | 3 |
| 5. | 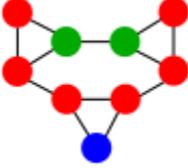 | (6, 2, 1) | 9 |

**Fig. 5.** Assembly pathways for a graph, a vector, and a number, which are homomorphic to each other. The vector's pathway is a vectorial addition chain. It is obtained by counting the number of nodes of each colour in the corresponding step of the graph's pathway. The number's pathway is an addition chain. It is obtained by counting all nodes in the graph in the corresponding step, or by summing the components of the vector in the corresponding step. Though the pathways for the graph and the vector are shortest pathways, the addition chain clearly is not, as step 4 is redundant.

As shown in [6], addition chain lengths have an easily calculable lower bound. Let s(n) denote the number of digits in the binary expansion of n. Then for any pathway P that constructs n:

$$l(P) \geq \log_2 n + \log_2 s(n) - 2.13$$

This bound thus serves as an easily calculable lower bound on the pathway assembly index of uniquely decomposable objects.

**Kolmogorov Complexity and Logical Depth**

In computer science Kolmogorov complexity [7] is one of the core complexity concepts, being the length of the shortest program (in a given Turing-complete language) that can produce a given output. Closely related is the notion of logical depth [8]. Given the shortest program that generates a given object, its logical depth is the computational time necessary to generate it from that program. A more involved definition is usually given, allowing the consideration of programs which are slightly longer than the shortest possible, but significantly faster to run, with the logical depth at a level of significance n considering all programs no more than n bits longer than the shortest, but we do not discuss this in detail here. In Bennett's this demonstrates an attempt to quantify the intuitive difference, in both physical and abstract cases, between "trivial" objects and those with a "nontrivial causal history" [8]. The same paper further considers its machine-independence and relation to self-organisation. Attempts have been made to apply logical depth and algorithmic complexity (of which Kolmogorov complexity is the most commonly used version) to biological systems and subsystems. Collier [9] proposes the use of logical depth as a general definition of "organisation" that can be applied to them. Other factors such as dynamic processes are considered relevant. Logical depth was specifically applied to simplistic ecological models [10] (Daisyworld models). They observed that the output of more detailed models had higher logical depth.

Although pathway assembly does not use Turing-complete languages as its main reference point, it shares the "shortest path" approach with these measures. With minor limitations imposed, the measure is computable, and so calculating it directly is practical. Conceptually, pathway assembly has a lot in common with these other minimisation based approaches, in particular with Kolmogorov complexity. Both pathway assembly and Kolmogorov complexity are concerned with finding the shortest means of generating an object. Neither, however, can be considered a special case of the other. It may be possible to use one to calculate bounds for the other indirectly, as the composition steps in pathway assembly are one of the many possible operations that a Turing-complete language can perform.

However, steps in pathway assembly potentially involve a lot more "free choice" per operation than a size-restricted section of a computer program, as there is no upper bound on the variety of objects that may be formed from any two particular objects in one step, and in many cases the combinatorial explosion of possible combinations would require more and more detailed programming instructions, with a corresponding longer length, to encapsulate them. That said, a correspondence may be possible in cases where this explosion does not occur, such as variants on addition chains, or the text example considered later in this paper. Likewise, logical depth is a measure of length that is dependent on another object (the computer program) for its own length (runtime).

**Algorithmic Information Theory and Effective Complexity**

Algorithmic information theory has been summarised [11] as, "an attempt to apply information-theoretic and probabilistic ideas to recursive function theory." It encompasses several core ideas in the study of complexity, such as the minimum number of bits required to specify an algorithm. It also considers probabilistic approaches to solutions. Coming at the topic from a cryptographic angle, one paper considers the minimal program and input necessary to generate a pseudo-random string, in the style of Kolmogorov complexity. The authors claim to demonstrate a strong relation between the incomputable Kolmogorov complexity and the computable measure known as Linear Complexity in all but a small number of cases for sufficiently large input. On the other hand, "effective complexity" is defined as a measure of the information content of the regularities of an object, originally introduced by Gell-Mann and Lloyd to avoid some of the perceived disadvantages of Kolmogorov complexity. It has been refined [12] with a precise definition in the language of algorithmic information theory, and analyse it further – for example, by comparing it to other measures. It also has a strong relation with logical depth. "If the effective complexity of a string exceeds a certain explicit threshold then that string must have astronomically large depth; otherwise, the depth can be arbitrarily small." Other attempts have been made to fully embody information theory concepts in physical processes, such as by starting from a physical definition of "meaningful information" [13].

**Applications**

**Text**

Pathway assembly has a straightforward application to strings of text, the most intuitive interpretation being to use the appropriate alphabet as the basis set and to let $f(a,b) = \{a +$

$b, b + a$}, where + represents the concatenation of strings from left to right, and where the basis set is an appropriate alphabet of single-character strings.

In the context of graphs described below, text can be considered to be a special case of directed graphs. If we allow reversed strings to be included in compositions (i.e. letting $x'$ denote the reverse of $x$, then $f(a,b) = \{a + b, b + a, a' + b, b + a', a + b', b' + a, a' + b', b' + a'\}$), then the system can be precisely modelled a coloured graph where each vertex has a degree no greater than 2. Disallowing this is also manageable, as described below.

## Graphs

Here we describe one method of representing graphs for pathway assembly purposes. We allow graph nodes to be coloured. Extra properties such directed edges are not discussed here, but easily included.

- $B = $ {a single-node graph of each colour}
- $f(a,b) = $ {all graphs G such that a and b are disjoint subgraphs of G, every vertex in G is in either a or b, and every edge in G is either in a, in b, or conn*ects* a vertex in a with a vertex in b}

Many contexts in which pathway assembly can be applied are easily reduced to graphs, and the algorithms we have developed are mainly designed with them in mind. Many structures, such as the molecules we consider in an upcoming paper, are most easily represented as graphs. The text example given above can be represented as a graph, with the extra stipulations that the graph's edges are directed, and that only graphs with a single directed pathway through all nodes, and no other edges, are permitted.

In the case of graphs, the divergence between Kolmogorov complexity and pathway assembly is particularly clear, as larger graphs have an ever-increasing number of ways to be combined with other large graphs, which can indefinitely be accomplished in a single step of an assembly pathway, but require a larger number of bits to specify in a computer program.

## Groups

In the mathematical field of group theory, a group is a pairing of a set G and an operation * that has the four following properties:

- Identity: $\exists e \in G$ such that $a * e = e * a = e \forall a \in G$. $e$ is called the identity.
- inverses: $\forall a \in G \ \exists \ a - 1 \in G$, called the inverse of $a$, such that $a * a - 1 = e$
- closure: $\forall a, b \in G, a * b \in G$

- transitivity: $\forall a, b, c \in G, (a * b) * c = a * (b * c)$

A generating set $B_G$ of a group $G$ is a set such that every member of $G$ can be expressed as a finite composition of members of $B_G$ and their inverses. The members of this set are called generators of the group.

From this, we can define the pathway assembly index of an element of a group like so:
- $U = G$
- $B = B_G$
- $f(a, b) = \{a * b, b * a, a^{-1} * b, b * a^{-1}, a * b^{-1}, b^{-1} * a, a^{-1} * b^{-1}, b^{-1} * a^{-1}\}$
  i.e. $f$ maps $a$ and $b$ to the set of all the ways that $a$ (or its inverse) and $b$ (or its inverse) can be composed within the group.

The members of any subset of $G$ have a co-assembly index similarly defined, and if $G$ is a finite group (i.e. has a finite number of members), we can define its pathway co-assembly index relative to $B$ and $*$ as the co-assembly index of all its members, although it is likely that objects in its pathway will not have all the properties of groups. Note that the pathway assembly of any of these objects is often highly dependent on the choice of generators.

The application of pathway assembly to groups may provide a useful foundation to applying it to objects that are not described with unique decompositions.

## Music

Many possible ways to represent music suggest themselves that pathway assembly can be applied to, depending which features of the music are to be considered. If music is to be represented by traditional notation (or many other representations that are discrete), the fundamental units could be absolute pitches, or the intervals between notes, the latter being more intuitively useful, as transposition of musical phrases is a common musical pattern, the symmetry of which would not be captured by the former. The duration of notes can also be incorporated.

One possible representation goes as follows:
- $B = \{\text{notes of various lengths, not bound to any specific pitch}\}$
- $f(a, b) = \{\text{a and b placed into relative temporal and pitch positions,}$
  $\qquad\qquad \text{preserving the relationships between all notes in a and in b}\}$

This allows for the introduction (and transposed repetition) of chords and of sections of melody. However, the musical structure within which a tune is considered can have a drastic effect on its assembly pathways.

Consider the well-known "Gloria" section of the Christmas Carol "Ding Dong Merrily On High" represented in figure 6. If restricted to a major scale, where the difference between a major or minor interval between two notes is determined by their position in the scale, then this music contains a single repeating section (in this case, a bar) five times over, lowering the pathway assembly index of this section. If the piece is instead considered within the context of a chromatic scale, this repetition is not exact, as the part is not a direct transposition of the notes each time.

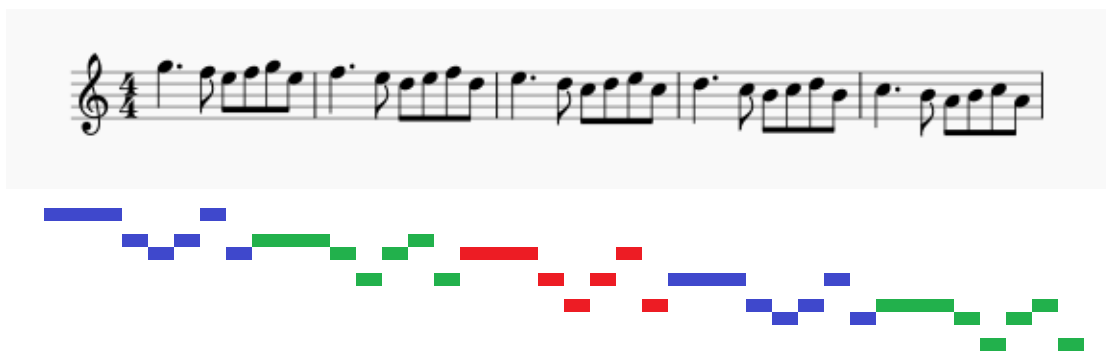

**Fig. 6.** Five bars of Ding Dong Merrily On High, written in the key of C Major, shown both in conventional musical notation and laid out chromatically. The pitch differences measured relative to steps of the major scale are always preserved, whereas the pitch differences measured relative to the frequency of the notes (assuming equal temperament) are only partially preserved, with three distinct patterns appearing.

## Image files

A grid of pixels (without the restriction that it must be rectangular) can be accurately modelled as a graph with coloured nodes. Alternatively, extra restrictions may be placed on the model, such as a limitation that orientation must be preserved, so that e.g. a section of the image and the same section rotated 90 degrees are not considered equivalent. In either case, nodes represent pixels, and nodes are connected if the pixels are neighbours. A simple repetitive image such as that in figure 7 would have a low assembly index.

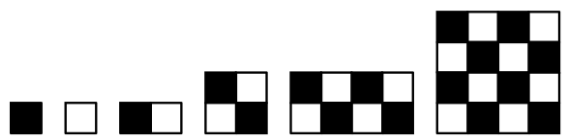

**Fig 7.** A chessboard pattern of pixels can be built up in few steps.

A typical bitmap image with many features and a high colour resolution, such as a photograph, is unlikely to have many repeating components. A simpler or monochrome image is more likely to have such components. The measure could also be applied not just to the image itself, but to associated structures, such as the quantisation matrices used in jpeg compression.

## Compression

Pathway assembly can be used to find space-efficient means of compression. For example, the Lempel-Ziv-Welch algorithm [14], one of the most well-established compression algorithms, can be adapted to take advantage of assembly pathways. The resulting algorithm is significantly slower than Lempel-Ziv-Welch during encoding, but can be decoded just as easily, and under favourable circumstances is more compressed than Lempel-Ziv-Welch. There are trade-offs between the two, depending on the input. This is further affected by arbitrary choices in the specifications of each algorithm, such as the amount of space allocated to symbols with different purposes, and so an objective comparison to find a clear "winner" is not feasible. It should also be noted that, as the compression methods discussed that are based on pathway assembly require a reasonable pathway to be found, we do not at this time have a method to complete the compression in a very short amount of time.

The algorithm below considers the compression for a text string in which all characters of the text are encoded with the numbers 0 to 255 under whatever encoding is use (e.g. unicode). The next "block" of 256 characters to specify the length of sub-objects, and the rest are used to represent stored blocks of text.

> 1. Find an assembly pathway for X. (This may be a shortest pathway if computing such a thing is feasible, but the algorithm will work with any pathway without redundancies.)
> 2. Identify all non-trivial objects on the pathway.
> 3. Let 512 be the next assignable code.
> 4. Identify the first occurrence within the string of a non-trivial object, disregarding occurrences that are not used on the pathway, and assign it the next assignable code. Increment the next assignable code by 1. (If two objects start at the same point, smaller objects have priority.)
> 5. Substitute all objects in the string that are used in the pathway with the symbol they have been assigned.
> 6. Replace the first occurrence of every stored symbol with the (compressed) text it represents, preceded by the symbol represented by 256 plus the length of that text.

This algorithm replaces all sections of text that appear on a pathway, and have some role in the pathway other than being incorporated once into a larger section of text, with the character that is used to represent them. The first instance of each, however, instead contains the text itself (or components that are used to build it), preceded by the number of characters that section of text contains. The assignment of sections of text to single characters can be deduced from the order in which they appear, allowing the text to be decompressed again.

## Conclusion

Pathway assembly's mathematical properties, even lightly explored, yield properties that assist its computation. It can be easily applied, at least cursorily, to several different fields, many of which may prove fruitful with further study. Furthermore, it provides a context within which abstractions of physical constructive processes can be studied, with which we intend to further explore the role of information in the formation of physical objects.